# Photonic nanobeam cavities with nano-pockets for efficient integration of fluorescent nanoparticles

*Johannes E. Fröch[†], Sejeong Kim[†,*], Connor Stewart[†], Xiaoxue Xu[‡], Ziqing Du[‡], Mark Lockrey[∥], Milos Toth[†], Igor Aharonovich[†,⁑]*

[†]School of Mathematical and Physical Sciences, [‡]Institute for Biomedical Materials and Devices, [∥]Microstructural Analysis Unit, University of Technology Sydney, NSW, 2007, Australia

[*]Sejeong.Kim-1@uts.edu.au   [⁑]igor.aharonovich@uts.edu.au

## ABSTRACT

**Integrating fluorescent nanoparticles with high-Q, small mode volume cavities is indispensable for nanophotonics and quantum technologies. To date, nanoparticles have largely been coupled to evanescent fields of cavity modes, which limits the strength of the interaction. Here, we developed both a cavity design and a fabrication method that enable efficient coupling between a fluorescent nanoparticle and a cavity optical mode. The design consists of a fishbone-shaped, one-dimensional photonic crystal cavity with a nano-pocket located at the electric field maximum of the fundamental optical mode. Furthermore, the presence of a nanoparticle inside the pocket reduces the mode volume substantially and induces subwavelength light confinement. Our approach opens exciting pathways to achieve strong light confinement around fluorescent nanoparticles for applications in energy, sensing, lasing and quantum technologies.**

Fluorescent nanoparticles have become indispensable elements in contemporary fields of science and research.[1-3] In the realm of quantum technologies and nanophotonics, there is great interest in increasing the emission rates from nanoparticles by modifying their photonic density of states via coupling to dielectric or plasmonic cavities.[4-14] Such hybrid systems can then be harnessed to realise low threshold lasers or utilised for quantum sensing and quantum cryptography applications.[15-22] Moreover, on-chip integration of nanoparticles with scalable photonic elements is essential for deployment of single photon sources in scalable quantum photonic applications.[23, 24]

The advantage of plasmonic cavities lie in their ease of coupling because of spectrally broad resonances. However, the devices are then inherently lossy due to metal absorption. Conversely, photonic crystal cavities (PCC) offer near-lossless operation, but integration is complicated due to spectrally narrow resonances. In both systems, nanoparticle placement accuracy is essential for efficient coupling to cavities. This is a significant challenge that has thus far not been resolved adequately. Pick-and-place techniques are time consuming and cumbersome;[25] two-step methods based on device fabrication and subsequent aligned-placement have low success probability and are applicable to a limited range of systems;[8, 14] and random dispersion of nanoparticles on cavities often results in unwanted residues that impede device performance.

Here we provide an elegant solution for hybrid integration of nanoparticles and dielectric cavities. We design and implement a one-dimensional (1D) PCC with a fishbone-shaped geometry and a nano-sized pocket for efficient integration of fluorescent nanoparticles, as shown in Figure 1a. This design overcomes problems encountered with common 1D PCCs (Figure 1b), which consist of a suspended beam and a series of airholes. In this case the altering refractive index between low (L) and high (H) index regions, give rise to a photonic bandgap, whilst modulation of the periodicity then introduces a photonic well at the center, confining light at the core of the nanobeam. Yet, a nanoparticle located on top of the cavity would only weakly interact with the optical mode formed in such a resonator. To increase this interaction a pocket can be introduced into the central H region of a PCC as indicated for the fishbone type cavity.[26] In the following, the rationale for the chosen geometry is given by two practical aspects. First, placement of nanoparticles by dispersion may easily clog airholes, hence introduce scattering and decrease the refractive index contrast along the propagation direction of the optical mode. This in turn can easily jeopardise the device functionality. In contrast, for a fishbone type cavity, (excess) nanoparticles can only attach to the sides of the nanobeam, thus disturbance due to nanoparticles for a propagating mode would only be present in the evanescent field, rendering this design more robust for practical application. Second, the footprint of a pocket relative to the fishbone structure is smaller as compared to airholes. This directly allows to exploit etching-lag in the same fabrication cycle to form a pocket, instead of etching a complete hole, as demonstrated in a later section. The structure allows not only for highly efficient coupling of emitters placed inside the pocket to the cavity, but the presence of a nanoparticle reduces the effective mode volume and increases the degree of light confinement.

Material of choice for the PCC is silicon nitride (refractive index *n=2*), due to its wafer scale availability and established fabrication recipes. The interaction with a nanoparticle is analysed using a photonic simulation tool (Lumerical Inc.) based on the finite-difference time-domain (FDTD) method. A nano-pocket with both diameter and depth of 80 nm is introduced in the middle of the fishbone cavity, where structural parameters are defined as follows: lattice constant ($a$)=230 nm, backbone width ($b$)=120 nm, bar width ($w_b$)=100 nm, bar length ($l_b$)= 500 nm, and thickness ($t$)=180 nm. To form a photonic well, the periodicity is then linearly decreased towards the center of the nanobeam. Electric field intensities for the cases with and without a nano-pocket are shown in Figure 1c and Figure 1d, respectively with a top (I), cross-sectional (II) and a magnified cross-sectional view (III). For comparison, the electric field intensities are normalized to the same scale. This comparison clearly shows that the optical mode and the nanoparticle interaction is stronger when the particle is placed in the pocket. The intensity at its center is 5.5 times higher than for a nanoparticle placed on top of the cavity. Moreover, nanosized gaps between particles and pocket sidewalls give rise to tight light confinement that results in a mode volume of $V \sim 0.015\ (\lambda/n)^3$. To the best of our knowledge, this is the smallest mode volume that was achieved for a dielectric cavity in the visible spectral range and comparable to the most efficient designs in the infra-red spectral range.[27-30] We note, that similar ultra-small mode volumes were previously achieved using plasmonic cavities, but these suffer from metallic losses, oxidation and challenges in assembly to typically sub 15 nm hot spots.

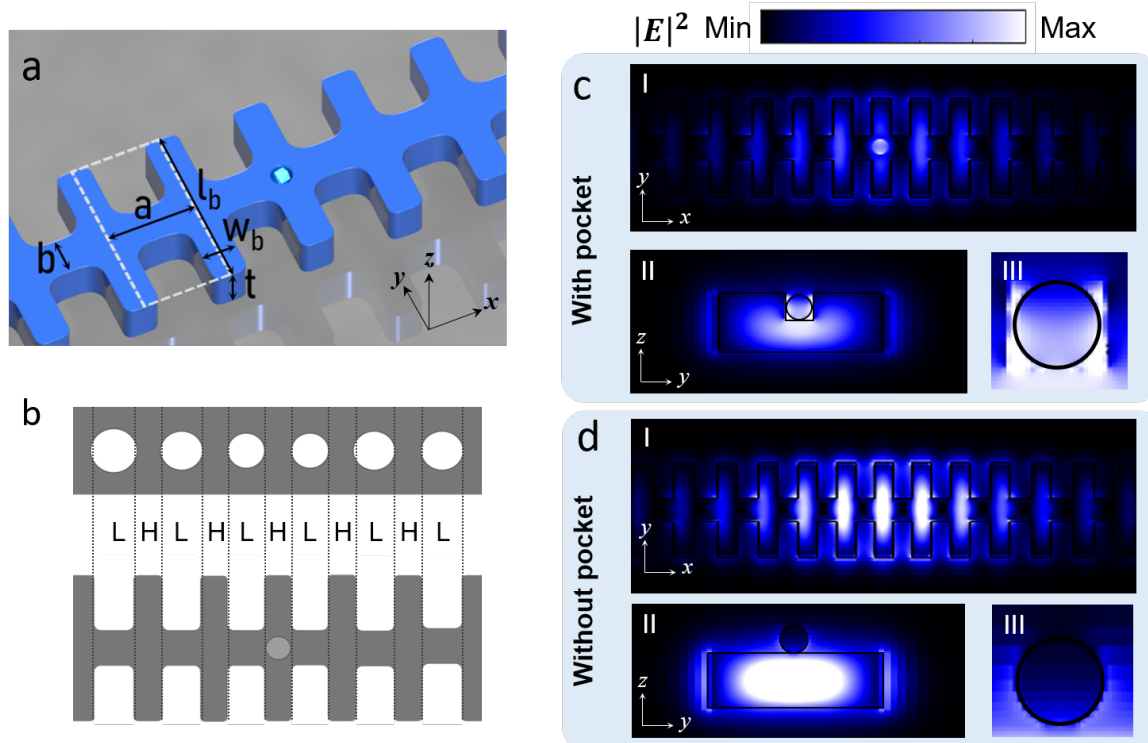

*Figure 1*. One-dimensional fishbone photonic cavity design. (a) 3D schematic of the proposed fishbone photonic crystal cavity (PCC) design with a nano-pocket placed in the middle of the cavity to host a fluorescent nanoparticle. The unit cell of the fishbone photonic crystal outlined by white dashed lines with periodicity a, backbone width b, bar length $l_b$, bar width $w_b$, and thickness t. (b) Refractive index modulation induced by altering regions of high (H) and low (L) refractive index either via airholes or by placing nanobars periodically on a backbone. FDTD simulation results showing the electric field intensity of (c) the cavity with nano-pocket and (d) without nano-pocket with (I) top-view, (II) cross-sectional view and (III) magnified view at the nanoparticle cross-section.

Following the conceptual design of the fishbone-pocket cavity we realised it experimentally, based on the fabrication steps outlined in Figure 2a (detailed in the Experimental Section). The process flow consists of (I) spin coating of the resist on a substrate, (II) electron beam lithography (EBL), (III) reactive ion etching (RIE) and immersion of the patterned substrate in a nanoparticle solution for particle placement, (IV) resist stripping and undercutting. A critical aspect of this work is the development of a single-cycle fabrication process, in which the cavity and the pocket are patterned and etched during the same EBL and RIE steps, respectively. This is possible, by taking advantage of RIE lag, where smaller features of the mask are etched at a slower rate, as compared to larger ones. Accordingly, a nano-pocket is automatically formed whilst the outline of the fishbone cavity is etched entirely. The described mechanism was observable by inspection in an SEM, as shown in Figure 2b for a set of free-standing PCCs after fabrication. In a magnified view (Figure 2c and 2d) of the highlighted areas it is clear that the outlines of device are clear defined and etched entirely. Moreover, we observed successful particle placement inside pockets as shown in Figure 2d. Furthermore, the characteristics of the pocket was directly confirmed by taking a cross section, shown in Figure 2e. Here a cross section along the center of the device clearly shows the formation of the intended pocket, which is etched approximately halfway through the slab. We emphasise that achieving different etching depths in a single RIE step is crucial and practical as scalable applications require minimum fabrication steps. In addition, it alleviates the stringent requirements of high alignment precision to create a pocket in a subsequent fabrication step for a pre-fabricated cavity.

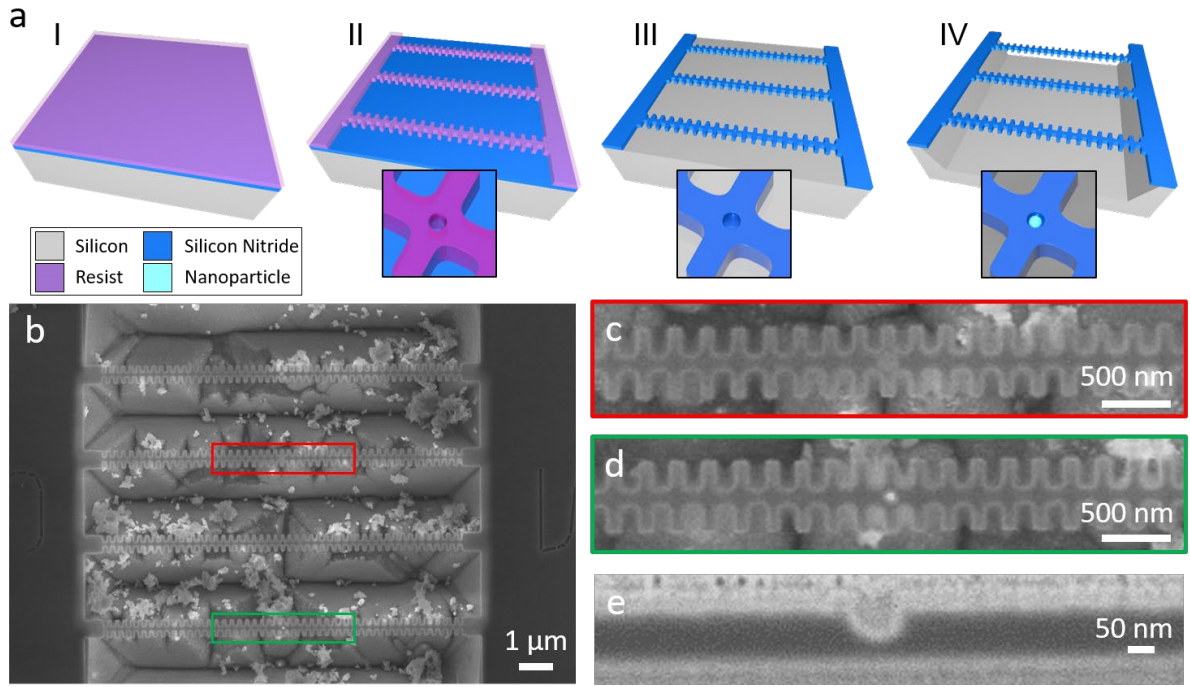

*Figure 2.* Fabrication of pocket cavities. (a) Schematic outline of the cavity fabrication scheme: (I) resist application on a SiN/ Si substrate, (II) patterning of the cavity outline and the nano-pocket in the center by EBL, (III) RIE placement in a nanoparticle solution (IV) final undercut of the cavity structure. (b) A set of PCCs after fabrication showing no nanoparticle contamination on the top of the cavity. Magnified SEM images showing cavities (c) without nanoparticle and (d) with nanoparticle in the nano-pocket, respectively. (e) Cross section of a pocket cavity showing the half-way through etched area of the nano-pocket due to the RIE lag effect.

In the next step we fabricated PCCs with resonances over a larger spectral region, 590 nm – 670 nm, a range relevant to a myriad of commonly deployed nanoparticle types (e.g. nanodiamonds, colloidal quantum dots, upconversion nanoparticles). Mode properties, device characteristics and the influence of the introduced pocket were then assessed using

Cathodoluminescence (CL) and Photoluminescence (PL), as specified in the Experimental Section. An SEM image of the fishbone cavity investigated for the conducted CL measurement is shown in Figure 3a. To reveal the spatial extent of the optical mode a CL spectral map was collected in the blue outlined area with a pixel size of 20 nm. The average collected spectrum in the entire region (blue) is shown in Figure 3b and compared to an averaged spectrum over the cavity center (red), showing a clear resonance at 595 nm. By integrating the CL signal from 593 nm to 597 nm an intensity map of the resonant mode could be reconstructed and correlated to the SEM image as shown in Figure 3c. This result matches well to the simulation, shown in Figure 1d, visualising the fundamental cavity mode with the maximum intensity at the center of the device.

Spectral device characteristics were then determined by PL measurements (532 nm excitation). A representative direct comparison of 2 PCCs, fabricated under identical fabrication conditions, with and without pocket, is shown in figure 3d. As could be expected, the cavity with a nano-pocket has a lower $Q$-factor ($Q$=1,100) compared to a cavity without pocket ($Q$=2,400) with a resonance blue-shifted by 3 nm due to a reduced effective refractive index. A statistical analysis of over 20 cavities with (red circles) and without pockets (blue crosses) is shown in Figure 3e, showing both tendencies over the entire range. Moreover, we observed an increasing trend of Q-factor with resonant wavelength for pocket-cavities, which is likely the result of reduced scattering from the nano-pockets.

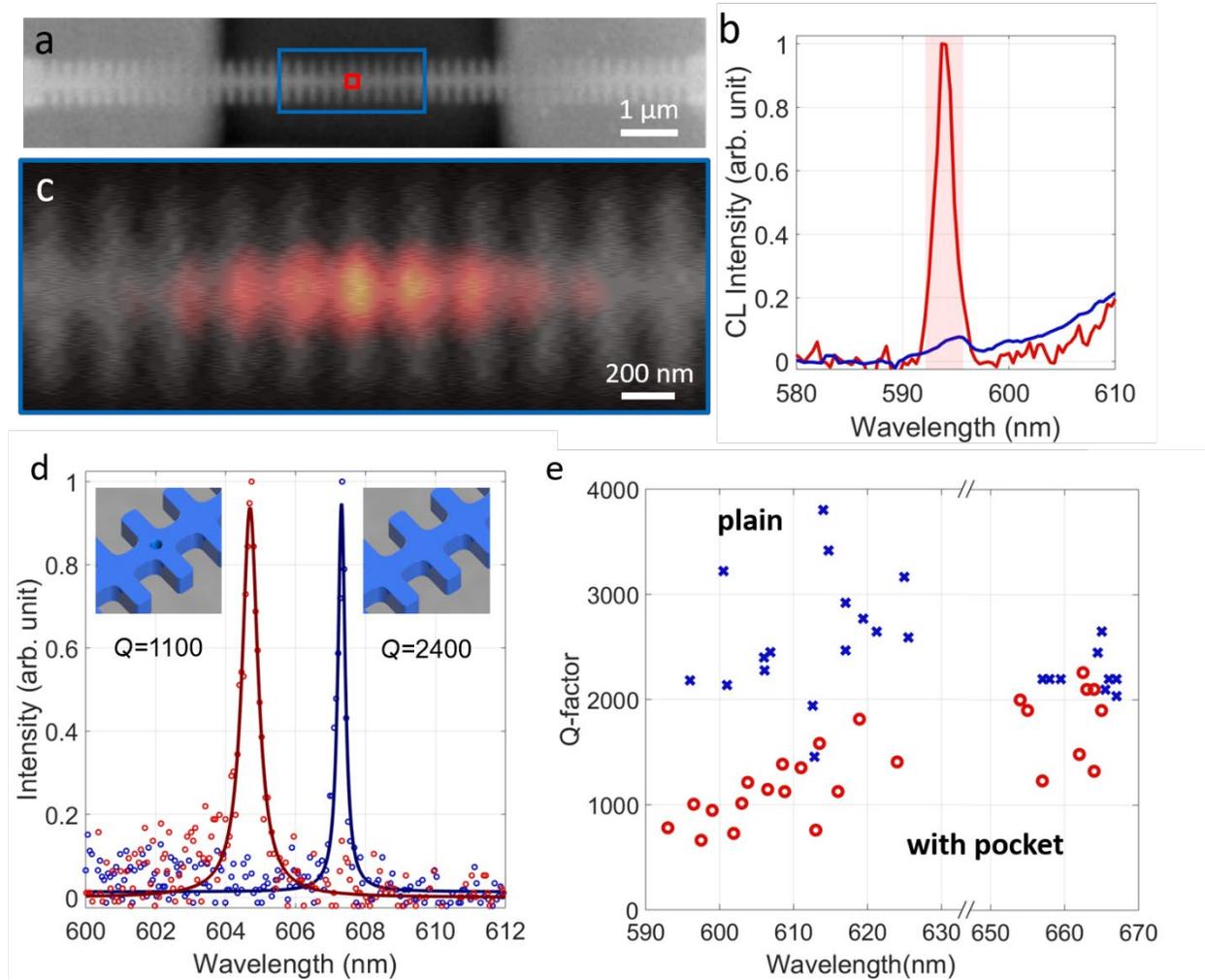

*Figure 3.* Spatial and spectral characterization of pocket cavities. (a) SEM image of the cavity investigated for the CL measurement. (b) Spectral response in CL of the structure averaged over the entire mapped area (red) and over the cavity center (blue), showing a resonant mode at 595 nm. (c) A 2D intensity map showing the spatial distribution of the optical mode at the cavity center. (d) Representative PL spectra showing cavity modes from PCCs with and without a nano-pocket. (e) Q-factors of all investigated PCCs in a range from 600 nm to 680 nm, determined from PL spectra (grating: 1200 lines/mm).

Next, we demonstrate the operation of the pocket cavity by enhancing the emission of a single upconversion nanoparticle (UCNP). Spherical NaYF$_4$ UCNPs (40 nm in diameter) co-doped with 48 % Yb$^{3+}$ and 2% Er$^{3+}$ were dispersed onto the cavities (during step IV, figure 2), yielding devices with nanoparticles as shown in Figure 4a. Some UCNPs are visible in between the bars of the fishbone cavity, but those do not degrade the experimentally measured $Q$ as shown below. Fluorescent properties were determined at room temperature by PL measurement under 980 nm laser excitation (cw, 300 µW). For the utilized pristine UCNP we observe a representative emission spectrum as shown in Figure 4b, with bands located at 521 nm, 544 nm, and 658 nm, corresponding to de-excitaiton of the Er$^{3+}$ after energy transfer from the primarily excited Yb$^{3+}$, shown schematically with the corresponding level diagram in Figure 4c. Targeting the red emission band at 658 nm (the $^4F_{9/2} \rightarrow {}^4I_{15/2}$ transition) we then observed coupled UCNPs to a cavity, as shown in Figure 4d. In comparison to the pristine emission (blue) the upconversion emission coupled to the cavity resonance at 665 nm with a $Q$-factor of 1,250 (inset) showed a 7-fold intensity enhancement. We note that only particles in the pocket contribute to the emission enhancement, as only their position overlaps with the spatial distribution of the optical mode (Figure 3c).

Further insight into the system was gained by lifetime measurements, pulsing the 980 nm laser. The collection was filtered using a bandpass filter and correlated to the pulsed excitation using a time correlator (Swabian Instruments). Background corrected lifetime measurements of pristine UCNPs (blue circles) and coupled to the cavity (red circles) are shown in Figure 4e, with the filtered spectrum shown in the inset. In case of pristine particles, the lifetime is best described by a single exponential fit yielding a value of 350 µs. To differentiate the lifetime of the coupled part, a double exponential fit was used, with the inverse spectral ratios and the lifetime of the uncoupled portion set as fixed input parameters. From the fit we then extracted a lifetime of 170 µs for the resonantly enhanced UCNP emission, thus giving a 2-fold lifetime reduction. We note here, that the presented factors for emission enhancement and lifetime reduction are in fact lower bounds. Because more particles are present in the excitation/collection spot it is expected that only a portion of the observed PL (i.e. from the particles in the pocket) was directly enhanced by the cavity.

Although the enhancement here demonstrated with UCNPs, the cavity design is very applicable to enhance quantum emitters in nanodiamonds, especially the emerging group IV defects that emit at the visible spectral range. Other immediate applications include nanoscale lasing with a vision of achieving single particle lasing that can be feasible with the current ultra small mode volume. Finally, we note that the cavity design is also suitable for single molecule spectroscopy that could be inserted into the cavity pockets.

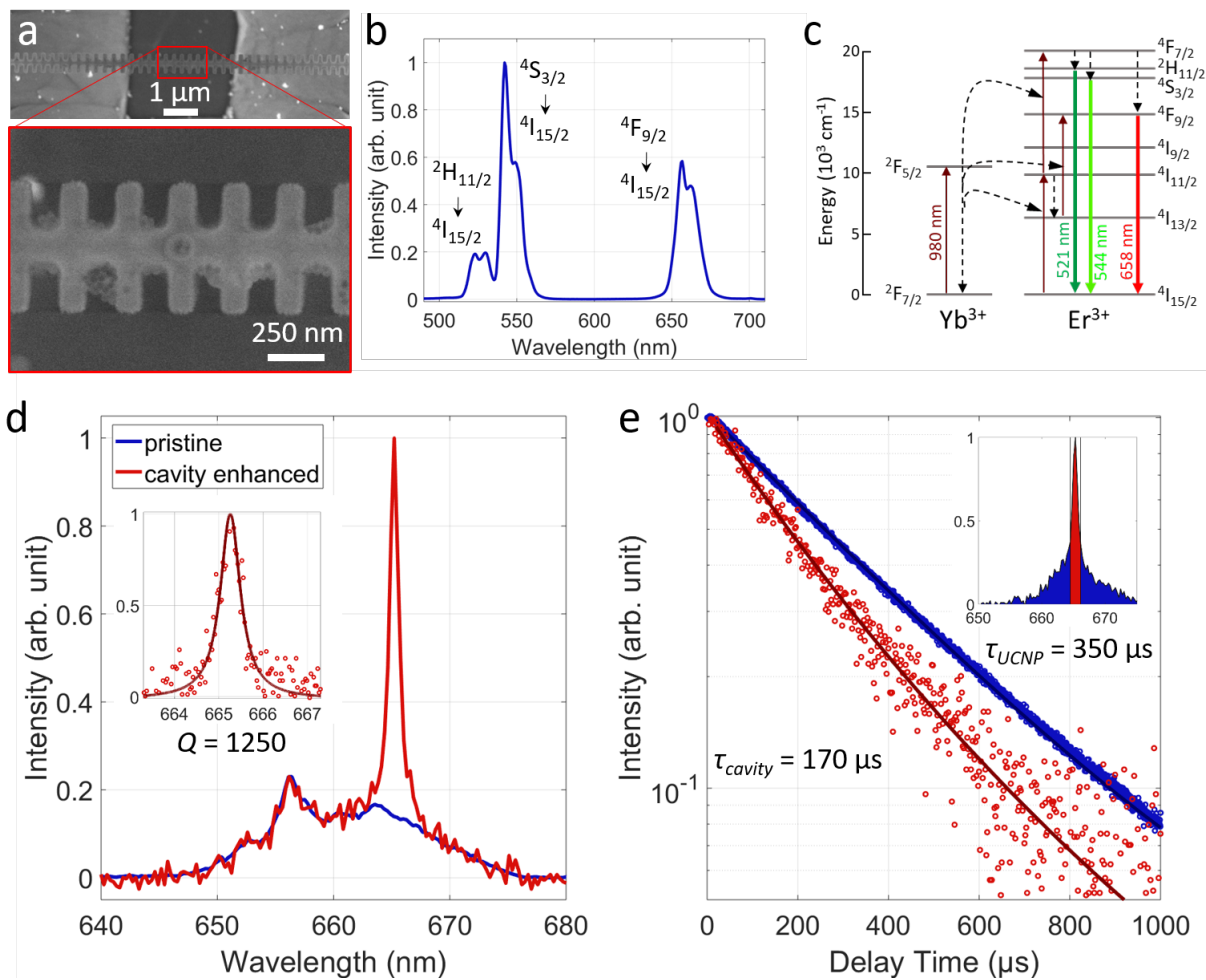

*Figure 4.* Fluorescent nanoparticles coupled to a pocket cavity. (a) SEM image of UCNPs inside a pocket of a cavity. (b) Representative emission spectrum of the utilized UCNP excited with a 980 nm laser with characteristic emission bands at 520 nm, 540 nm, and 660 nm. (c) Level diagram for the upconversion excitation process of the $Er^{3+}$ dopant. (d) Particles coupled to a cavity (red) exhibit a 7-fold emission enhancement at 665 nm at the cavity resonance compared to the typical emission at this wavelength (blue). The cavity has a Q-factor of 1,250 (inset). (e) Lifetime measurement under pulsed excitation of pristine UCNPs (blue) and the coupled system (red). The measurement for the uncoupled UCNP was fitted to a single exponential (dark blue) yielding a lifetime of 350 µs. The lifetime of the coupled part was extracted by a double exponential, using

*the PL area ratios (inset) and the lifetime of the uncoupled UCNPs (350 µs) as fixed input parameters for the fit, yielding a lifetime of 170 µs.*

To summarize, we proposed and realised novel 1D fishbone cavities with a nanoscale pocket for fluorescent nanoparticles. The presence of a nanoparticle inside the pocket reduces the mode volume significantly leading to a calculated modal volume of 0.015 $(\lambda/n)^3$. The cavities are fabricated using a single step EBL-RIE and the cavities exhibit high quality factors exceeding $1\times10^3$. As a proof of concept, we ultimately demonstrated coupling of a UCNP to a pocket cavity, where we achieved a 7-fold PL intensity enhancement. Overall, the demonstrated design is universally and immediately applicable to all types of fluorescent nanoparticles, such as nanodiamonds, UCNPs, or colloidal quantum dots, and tailored towards applications in nanoscale lasers quantum photonics and sensing.

**Experimental Section**

**Device Fabrication**. A detailed description of the fabrication steps, given in Figure 2a are as follow: (I) a resist (CSAR 62 11 % - ALLRESIST GmbH.) was spun (500 nm) onto a $Si_3N_4$ (180 nm) on Si substrate and pre-baked at 160 °C for 1 minute. Then, (II) EBL (Zeiss Supra 55 VP and RAITH EBL system) was used to define the cavity pattern including a hole with an 80 nm diameter at the center of the nanobeam. After developing the E-beam resist, the pattern was transferred into the underlying substrate using RIE (90 W, 300 V self-bias, 60 sccm, 10 mTorr) with $SF_6$ as the reactive gas. After RIE, the substrate was immersed in a solution of nanoparticles. During solvent evaporation, nanoparticles are assembled inside pockets. The resist film was then stripped off in a heated acetone bath which removes the remaining resist and excess nanoparticles other than those located in the nano-pockets. In a final step the nanobeams were undercut, using a 20 wt. % KOH solution (40 °C, ~ 20 min.) etching the silicon underneath.

**CL Characterisation.** The device presented in figure 3a was analysed in a FEI DB235 equipped with a DELMIC SPARC CL system operated at an acceleration voltage of 10 kV and beam current of 4 nA. Acquisition time per pixel was set to 500 ms, and dispersed by a 1800 lines/mm grating with the entry slit opened to 2 mm to maximise the collected signal.

**PL Characterisation.** Photoluminescence characterization of PCCs was carried out using a confocal micro PL setup with a 532 nm laser (continuous wave, 500 µW at the objective back focal plane) as the excitation source and collection through a 0.9 NA objective. For spectral acquisition a grating of 1200 lines/mm was used. For characterisation of UCNPs we used a 980 nm laser as the excitation source (300 µW, at the objective back focal plane). Lifetime measurements were conducted by directing the laser through an acousto-optic modulator (AOM), setting the period to 500 µs pulse width/ 1500 µs delay. After the AOM, the 1st order of the diffracted laser was used as the excitation source (300 µW, at the objective back focal plane). The signal was then coupled into a fiber and guided to an APD (excelitas).

**Author contributions**

**Acknowledgement**

The authors thank the Australian Research Council (DP180100070, DP190101058) and the Office of Naval Research Global under grant number N62909-18-1-2025 for financial support. Authors gratefully thank Dr. Carlo Bradac and Dr. Mehran Kianina for help with PL measurements, setting up of the lifetime measurements and useful discussions, as well as Dr. Fan Wang for supporting us with a 980 nm laser.